# On the polynomial depth of various sets of random strings

Philippe Moser *

**Abstract**

This paper proposes new notions of polynomial depth (called monotone poly depth), based on a polynomial version of monotone Kolmogorov complexity. We show that monotone poly depth satisfies all desirable properties of depth notions i.e., both trivial and random sequences are not monotone poly deep, monotone poly depth satisfies the slow growth law i.e., no simple process can transform a non deep sequence into a deep one, and monotone poly deep sequences exist (unconditionally).

We give two natural examples of deep sets, by showing that both the set of Levin-random strings and the set of Kolmogorov random strings are monotone poly deep.

## 1 Introduction

From the observation that nature contains both very simple and highly complex structures, Bennett introduced the profound concept of logical depth [7], as a formal definition of *useful* information, as opposed to (random) information in the traditional algorithmic information theory sense. Bennett's original idea is to categorize structures in three groups: trivial, random and the remaining ones; with the idea that trivial structures being completely predictable contain no useful information; random ones, being completely unpredictable, do not contain any useful information either; both (trivial and random) being therefore shallow objects. On the other hand, structures that are neither random nor trivial i.e., that contain intricate patterns that are neither fully predictable nor completely unpredictable, contain useful information; they are called deep structures. Although random sequences contain a lot of information (in the sense of algorithmic information theory), this information is not of much value, and such sequences are shallow.

Bennett observed that deep objects, because they contain complex well-hidden patterns, cannot be created by easy processes. This observation was formalized in the so-called *slow growth law*, which states that if a simple process (a truth table reduction) transforms some (source) sequence into an (image) sequence that is deep, then the source sequence it started from must be deep i.e., no easy process can transform a shallow sequence into a deep one.

Bennett's logical depth is based on Kolmogorov complexity. Intuitively, a binary sequence is deep, if the more time an algorithm is given, the better it can compress the sequence. Although Bennett's formulation is theoretically very elegant, it is uncomputable, due the uncomputability of Kolmogorov complexity.

To overcome the uncomputability of logical depth, several notions of feasible depth have been proposed so far [11, 5, 9]. In [5] Antunes, Fortnow, van Melkebeek, and Vinodchandran studied several polynomial-time formulations of depth, with connections to average-case

*Department of Computer Science, National University of Ireland Maynooth, Maynooth, Co. Kildare, Ireland.

complexity, nonuniform circuit complexity, and efficient search for satisfying assignments to Boolean formulas. In [9], both a notion of finite-state and polynomial depth were investigated, and the depth of polynomial weakly useful languages was shown.

Unfortunately, the feasible notions proposed so far suffer some limitations, e.g. a notion in [5] requires a complexity assumptions to prove the existence of deep sequences; and the polynomial depth of [9] is based on polynomial time predictors that cannot read their input (predictors must predict the $n$th bit of a sequence without access to the history, i.e. bits $1, 2, \ldots, n-1$).

As noticed in [15], depth is not an absolute concept, but depends on the power of two competing group of observers $\Delta$ and $\Delta'$. Informally a sequence is $(\Delta, \Delta')$-deep if for any observer $O$ from $\Delta$ there is an observer $O'$ in $\Delta'$ such that $O'$ performs (e.g. compresses, predicts, etc.) better than $O'$ on the sequence. $\Delta$ and $\Delta'$ can be the same class e.g. for recursive depth [11], $\Delta = \Delta'$ are recursive time bounds, or different classes e.g. for Bennett's depth [7], $\Delta$ are recursive time bounds but $\Delta'$ is unbounded Kolmogorov complexity.

In this paper, we use the idea of competing observers from [15] to construct new notions of polynomial depth (called monotone-polynomial depth), aiming at notions that satisfy the slow growth law, and for which deep objects can be proved to exist unconditionally. The classes of observers (the classes $\Delta$ and $\Delta'$) we consider are based on the notion of monotone polynomial time compression [8], which is a polynomial version of monotone Kolmogorov complexity, with the advantage that unlike polynomial predictors [9], monotone polynomial compressors can read their input. We show that our notions of monotone polynomial depth have all the desired properties of a depth notion, i.e. both trivial and random sequences are shallow, they satisfy a slow-growth law, and deep objects can be shown to exist unconditionally.

Although logical depth is a very profound concept, there have not been many examples of natural deep sequences in the literature so far. Bennett [7] showed that the halting language is deep. Lathrop and Lutz [10] generalized Bennett's result by showing that every weakly useful sequence (i.e. a sequence such that the set of languages that can be reduced to it has measure non-zero) is deep, a result that was shown to hold in the context of polynomial depth [9]. In this paper, we give two natural examples of deep languages, in the context of monotone poly depth, namely the set of Levin-random strings and the set of Kolmogorov random strings. Levin randomness is a standard randomness notion due to Levin [12]; it is a computable approximation of Kolmogorov complexity, that enjoys many useful properties, among others it provides a search strategy for finding solutions of NP problems, that is optimal up to a multiplicative constant (see [13]). Curiously although random sequences are shallow, our result shows that the set of Levin-random strings is not. This shows that in the context of polynomial monotone depth, having a test that detects randomness (i.e the set of Levin-random strings), is more useful than having access to randomness (a random sequence).

Several authors [2, 4, 1, 3] showed the computational power of the set of Kolmogorov random strings by reducing (using several types of reduction) a broad range of complexity classes to it. Our observation that the set of Kolmogorov random strings is monotone-poly deep is consistent with the results by these authors [2, 4, 1, 3] whose results intuitively show that this set contains a lot of useful information.

Due to lack of space, some proofs are postponed to the appendix.

## 2 Preliminaries

We write $\mathbb{N}$ for the set of all nonnegative integers. Let us fix some notations for strings and languages. A *string* is an element of $\{0,1\}^n$ for some integer $n$. We denote by $s_0, s_1, \ldots, s_n$ the standard enumeration of strings in lexicographic order. For a string $x$, its length is denoted by $|x|$. The empty string is denoted by $\lambda$. We say string $y$ is a prefix of string $x$, denoted $y \sqsubset x$ (also $y \sqsubseteq x$), if there exists a string $a$ such that $x = ya$. We write $x \sim y$ if $x$ is a prefix of $y$ or vice-versa. For a string $x$, dbl($x$) is $x$ with every bit doubled.

A sequence is an infinite binary string, i.e. an element of $\{0,1\}^\infty$. For $S \in \{0,1\}^\infty$ and $i,j \in \mathbb{N}$, we write $S[i..j]$ for the string consisting of the $i^{\text{th}}$ through $j^{\text{th}}$ bits of $S$, with the convention that $S[i..j] = \lambda$ if $i > j$, and $S[1]$ is the leftmost bit of $S$. We write $S[i]$ for $S[i..i]$ (the $i^{\text{th}}$ bit of $S$). For a sequence $S$ divided into blocks $S = S_1 S_2 S_3 \ldots$, where $S_i$ are strings, $S \upharpoonright S_i$ (resp $S \upharpoonleft S_i$) denotes $S_1 \ldots S_i$ (resp. $S_1 \ldots S_{i-1}$). For $w \in \{0,1\}^*$ and $S \in \{0,1\}^\infty$, we write $w \sqsubseteq S$ if $w$ is a prefix of $S$, i.e., if $w = S[1..|w|]$. Unless otherwise specified, logarithms are taken in base 2.

A *language* is a set of strings. The characteristic sequence of a language $L$ is the sequence $\chi_L \in \{0,1\}^\infty$, whose $n$th bit is one iff $s_n \in L$. We will often use the notation $L$ for $\chi_L$.

TM stands for Turing machine. A monotone TM is a TM such that for any strings $x, y$, $M(xy) \sqsupseteq M(x)$.

E denotes the standard linear exponential time complexity class $\mathsf{E} = \cup_{c \in \mathbb{N}} \mathsf{DTIME}(2^{cn})$. A time bound is a monotone time constructible function $t : \mathbb{N} \to \mathbb{N}$, i.e. there is a TM that on input any string of length $n$ halts in exactly $t(n)$ steps. We will consider the following standard time bound families: Poly $= \cup_{k \in \mathbb{N}} \{t(n) = kn^k\}$, Lin $= \cup_{k \in \mathbb{N}} \{t(n) = kn\}$, Polylog $= \cup_{k \in \mathbb{N}} \{t(n) = \log^k n\}$ and Rec $= \{t|\ t$ is a time bound$\}$.

## 3 Polynomial depth

Our polynomial depth notions are based on polynomial monotone compression from [8].

**Definition 3.1** *Let $\Delta$ be a family of (at least linear) time bounds (e.g. Poly, Lin, etc) and $S \in \{0,1\}^\infty$. A $\Delta$-compression of $S$ is a 3-tuple $(C, D, p)$ where $C, D$ are TMs and $p \in \{0,1\}^\infty$ such that there exists a time bound $t \in \Delta$ such that*

1. *Decompression: For all $j \in \mathbb{N}$, $D(p[1..j])$ outputs $S[1..i_{D,j}]$ in time $t(i+j)$, where $i_{D,j}$ is a monotone sequence of integers.*

2. *Compression: For all $i \in \mathbb{N}$, $C(S[1..i])$ outputs several strings in time $t(i)$, one of which is a prefix $p'$ of $p$, such that $D(p') \sqsupseteq S[1..i]$.*

The integer $i_{D,j}$ is the number of bits the decompressor $D$ can output given $i$ bits of input i.e., the larger the difference $i_{D,j} - j$ the greater the compression.

When $\Delta = $ Rec, we drop the compression requirement, i.e. a Rec-compression is a 2-tuple $(D, p)$. This is because the compressor $C$ may be uncomputable. When $\Delta = $ Rec we are in the realms of Kolmogorov complexity, where similarly there is no (computable) compressor but only a computable decompressor (the universal TM $U$).

To avoid extreme compressions of the form "On input $n$, output $2^{2^{\cdot^{\cdot^{\cdot^{2^n}}}}}$ zeroes", we fix the maximal compression factor we allow i.e., let MD (maximal decompression) be a function such that MD($j$) is computable in $O(\mathrm{MD(j)})$ time for any integer $j$ (e.g. MD($j$) = $2^{2^{2^j}}$). We require that for any $\Delta$-compression $(C, D, p)$, and for every integer $j$,

$$i_{D,j} \leq \mathrm{MD}(j)$$

i.e. MD is the same for all compressors.

Let us introduce our $(\Delta, \Delta')$-depth notion, based on competing observers' classes $\Delta$ and $\Delta'$.

**Definition 3.2** *$S \in \{0,1\}^\infty$ is a.e. (resp i.o.) $(\Delta, \Delta')$-deep if for every $\Delta$-compression $(C, D, p)$ of $S$ and any $a > 0$, there exists a $\Delta'$-compression (C',D',p') of $S$ such that for almost every (resp. infinitely many) $j \in \mathbb{N}$*

$$i_{D',j} - i_{D,j} \geq a \log i_{D',j}. \tag{1}$$

A sequence is $(\Delta, \Delta')$-shallow if it is not $(\Delta, \Delta')$-deep.

The choice of the log function in Equation 1 is arbitrary. In Bennett's original notion [7], it was only required the difference be unbounded and the rate was not specified, but Bennett's notion would also work with a log rate function. Most feasible depth notions published after Bennett's paper [5, 9] used a logarithmic rate function. We choose to do the same.

As noticed in [15], the notion of depth is a relative notion, that depends on the power of the observers. Our goal is to study polynomial versions of Bennett's original depth notion [7], and its recursive version called recursive depth [11]. Recursive depth [11] is defined in terms of recursive observers competing against recursive observers, i.e. $\Delta$ and $\Delta'$ have the same power. The natural polynomial version is to choose $\Delta = \Delta' = $ Poly. We call this notion monotone-Poly-depth.

Bennett's depth [7] on the other hand is based on observers of different strength i.e., recursive observers competing against noncomputable Kolmogorov complexity. For Bennett's notion, there is no unique translation into the polynomial world. We propose to study ($\Delta = $ Lin, $\Delta' = $ Poly) as a polynomial version of Bennett's depth (called monotone-Lin-depth), which encompasses the idea of observers of different strength (Lin vs Poly), but keeping both in the polynomial setting. The choice ($\Delta = $ Lin, $\Delta' = $ Poly) is actually flexible, and Poly (resp. Lin) could be replaced by anything strictly stronger (resp. weaker ) than Lin, e.g. $O(n^2)$ (resp. Polylog), without modifying our results on monotone-Lin-depth from Section 5 (the choice $\Delta = $Polylog, would require a modification of the notion of $\Delta$-compression [8] to allow for sublinear running time, in the same way as martingales where modified to allow sublinear time bounds in [16]. We defer this generalization to the full version of this paper). The choice Lin vs Poly somehow reflects the difference in power of complexity classes E and EXP, which are the complexity classes on which $\Delta$-compression was first introduced [8], to define a measure notion.

In [5] Antunes et al. proposed another resource-bounded version of Bennett's depth [7] called basic computational depth, by looking at bounded (sublinear or polynomial) Kolmogorov complexity vs unbounded Kolmogorov complexity. We introduce a translation of basic computational depth [5] in the setting of polynomial monotone compressors, by choosing ($\Delta = $ Poly, $\Delta' = $ Rec). We call this notion basic-monotone-Poly-depth (bm-Poly-depth). bm-Poly-depth captures the idea behind basic computational depth [5] but with Kolmogorov complexity replaced by monotone compressors.

The difference between a.e. and i.o. depth is similar to the difference between (resource-bounded)-packing dimension and (resource-bounded)-dimension (see e.g. [6]), where a compressor is required to compress infinitely many prefixes, or almost all prefixes. Bennett's depth [7] is an a.e. notion. Sometimes when the observers are very weak e.g. finite-state, i.o. is the best achievable (e.g. see [9]). All our results use the stronger formulation i.e. a.e. (which implies an i.o. result), except Theorem 4.4.

## 4 Basic properties of monotone-Poly-depth

In the next section we study the basic properties of monotone-Poly-depth. All results remain true for both monotone-Lin-depth and bm-Poly-depth.

It is a key feature of logical depth [7] that both trivial (recursive) and random sequences are shallow. In this section we show that a similar result holds in the context of monotone-Poly-depth. Let us define what is meant by trivial sequences in the context of polynomial depth. Informally a sequence is trivial if its prefixes can be maximally compressed.

**Definition 4.1** *Let $S \in \{0,1\}^\infty$. $S$ is Poly-optimally-compressible if there exists a Poly-compression $(C, D, p)$ of $S$, such that $i_{D,j} = MD(j)$ for almost every $j \in \mathbb{N}$.*

As an example, it is easy to check that the characteristic sequences of languages in E are Poly-optimally-compressible. The following result shows that optimally-compressible sequences are shallow.

**Theorem 4.1** *Every Poly-optimally-compressible sequence is a.e. Poly-shallow.*

On the other extremity of the scale of randomness, we have random sequences. Here is a definition in the context of polynomial depth.

**Definition 4.2** *Let $S \in \{0,1\}^\infty$. $S$ is Poly-random if for every Poly-compression $(C, D, p)$ of $S$, there exists $c \in \mathbb{N}$ such that for almost every $j \in \mathbb{N}$*

$$i_{D,j} \leq j + c.$$

The following result shows that random sequences are shallow.

**Theorem 4.2** *Every Poly-random sequence is a.e. Poly-shallow.*

### 4.1 Slow growth law

A key property of logical depth [7], is that depth cannot be easily created. The formalization of this idea is known as the slow-growth law. It states that if a simple process transforms some (source) sequence into an (image) sequence that is deep, then the source sequence it started from must be deep i.e., no easy process can transform a shallow sequence into a deep one. Bennett proved a slow growth law for truth-table reductions (i.e. in the context of logical depth, simple process corresponds to truth-table reductions).

In the following section, we prove a slow growth law in the context of monotone-Poly-depth. As the power of polynomial monotone compressors is much smaller than the unbounded time case considered for Bennett's logical depth, we need to reduce the power of "simple processes" accordingly, by choosing weaker reductions. Here is a definition.

**Definition 4.3** *Let $S, T \in \{0,1\}^\infty$. $S$ is Poly-monotone reducible to $T$, if there exists a Poly-time monotone TM $M$ such that*

1. *Reduction: for every $n \in \mathbb{N}$, $M(T[1..n]) \sqsubseteq S$.*

2. *Honesty: There exists $a > 0$ such that for every $n \in \mathbb{N}$*

$$n - a \log n \leq |M(T[1..n])| \leq n + a \log n$$

3. *Monotone injectivity: If $M(x) \sim M(y)$ then $x \sim y$.*

The following result is a slow-growth law for monotone-Poly-depth. A similar result holds for both monotone-Lin-depth and bm-Poly-depth (provided the reduction is linear-time bounded for monotone-Lin-depth).

**Theorem 4.3** *Let $S, T \in \{0,1\}^\infty$, such that $S$ is a.e. monotone-Poly-deep and Poly-monotone reducible to $T$. Then $T$ is a.e. Poly-deep.*

A similar proof shows that the result holds for both monotone-Lin-depth and bm-Poly-depth (provided the reduction is linear-time bounded for monotone-Lin-depth).

## 4.2 A Poly deep sequence

Some previous polynomial depth notions (e.g. distinguishing complexity from [5]) require complexity assumption to prove the existence of deep sequences. The following result shows that our notion is unconditional. Similarly to other feasible depth notions with restricted power [9], our result is an i.o. result.

The proof uses the equivalence between compressors and martingales from [8]. A direct proof can be given without martingales, but using martingales makes the proof easier to read since it is easier to sum and diagonalize against martingales than it is against compressors directly. It is also interesting to see the correspondence martingales-compressor in the context of depth.

**Theorem 4.4** *There exists an i.o. monotone-Poly-deep sequence.*

## 5 The set of Levin random strings is deep

Whereas random sequences are shallow, we show that the characteristic sequence of the set of random strings is deep. Our result holds for the standard randomness notion due to Levin [12]; Levin's notion is a computable approximation of Kolmogorov complexity, that enjoys many useful properties, among others it provides a search strategy for finding solutions of NP problems, that is optimal up to a multiplicative constant (see [13]). Here is a definition.

**Definition 5.1** *Fix a universal Turing machine $U$. The Levin complexity of a string $x$ is*

$$\mathrm{Kt}(x) = \min\{|p| + \log t :\ U(p) = x \ \text{in at most } t \ \text{steps}\}.$$

The definition of Kt does not depend on the choice of the universal TM $U$, up to an additive constant (see [13]).

The set of Levin random strings is

$$R_{\text{Kt}} = \{x \in \{0,1\}^* : \text{Kt}(x) \geq |x| + \log|x|\}. \tag{2}$$

By a standard program counting argument, it is easy to see that $R_{\text{Kt}} \neq \emptyset$. Although the strings in $R_{\text{Kt}}$ are shallow, the characteristic sequence of $R_{\text{Kt}}$ contains useful information, i.e. is monotone-Lin-deep, as the following result shows.

**Theorem 5.1** *$R_{\text{Kt}}$ is a.e. monotone-Lin-deep.*

*Proof.* We need the following lemma.

**Lemma 5.1** *For every Lin-compression $(C, D, p)$ and for almost every $j \in \mathbb{N}$*

$$i_{D,j} < 2^{2^{3j}+1}.$$

Let us prove the lemma by contradiction. Suppose there is an infinite set $J$ of integers $j$ such that $i_{D,j} \geq 2^{2^{3j}+1}$; in particular for every $j \in J$, $i_{D,\frac{1}{3}\log j} > 2^{j+1}$. Thus

$$R_{\text{Kt}} \sqsupset D(p[1..\frac{1}{3}\log j]) \sqsupset R_{\text{Kt}}[1..2^{j+1}].$$

Let $j \in J$ be large (to be determined later). Letting $d = p[1..\frac{1}{3}\log j]$ yields a string with high Kt complexity: from $\pi = \langle D, d \rangle$ recover $R_{\text{Kt}}[1..2^{j+1}]$ and $j$ (from the length of $d$). Output the first $y$ with $|y| = j$ and $R_{\text{Kt}}(y) = 1$, i.e.

$$\text{Kt}(y) \geq j + \log j.$$

By encoding $\pi$ the standard way, i.e. $\pi = \text{dbl}(\langle D \rangle)01d$

$$|\pi| \leq |d| + O(1) \leq \frac{1}{3}\log j + O(1).$$

The time to construct $j$ is the time to recover $R_{\text{Kt}}[1..2^{j+1}]$ (less than $O(2^{j+1})$) and the time to find $y$ in $R_{\text{Kt}}[1..2^{j+1}]$ (less than $O(2^{j+1})$ steps), i.e. a total of at most $O(2^{j+1})$ steps. Therefore

$$\text{Kt}(y) \leq |\pi| + \log O(2^{j+1}) \leq \frac{1}{3}\log j + j + O(1) < j + \log j$$

for $j$ large enough, which contradicts $R_{\text{Kt}}(y) = 1$; thus ending the proof of the lemma. □

**Lemma 5.2** *There exists a Poly-compression $(C, D, p)$ of $R_{\text{Kt}}$ such that for almost every $j \in \mathbb{N}$*

$$i_{D,j} = \text{MD}(j).$$

Let $p = 0^\infty$. $D$ on input $p[1..j]$ computes $i_{D,j} := \text{MD}(j)$. $D$ constructs $R_{\text{Kt}}[1..i_{D,j}]$ by simulating the universal TM on all programs $\pi_l$ of size at most $\log i_{D,j} + \log \log i_{D,j}$ during $t_l$ steps ($t_l \leq 2^{\log i_{D,j} + \log \log i_{D,j}}$), the results string of such a simulation is denoted $x_l$. All strings $x_l$ with $|x_l| \leq \log i_{D,j}$, for which $|\pi_l| + \log t_l \leq |x_l| + \log|x_l|$ have membership bit 0 in the characteristic sequence $R_{\mathrm{Kt}}[1..i_{D,j}]$. All remaining bits in $R_{\mathrm{Kt}}[1..i_{D,j}]$ are 1s. The running time of $D$ is less than

$$O(2^{\log i_{D,j} + \log\log i_{D,j}}) \cdot 2^{\log i_{D,j} + \log\log i_{D,j}} \leq (i_{D,j})^c$$

for some $c \in \mathbb{N}$.

The compressor $C$ on input $R_{\mathrm{Kt}}[1..i]$ finds the smallest $j$ such that $\mathrm{MD}(j) \geq i$, and outputs $0^j$. $C$ runs in time polynomial in $i$. This ends the proof of the lemma. $\square$

Let us show that $R_{\mathrm{Kt}}$ is monotone-Lin-deep. Let $a > 0$ and $(C, D, p)$ be a Lin-compression of $R_{\mathrm{Kt}}$, and let $(C', D', p')$ be the Poly-compression from Lemma 5.2. We have

$$\begin{aligned}
i_{D',j} - i_{D,j} &= \mathrm{MD}(j) - i_{D,j} && \text{by Lemma 5.2} \\
&\geq \mathrm{MD}(j) - 2^{2^{3j}+1} && \text{by Lemma 5.2} \\
&\geq \frac{1}{2}\mathrm{MD}(j) && \text{by definition of MD} \\
&= \frac{1}{2} i_{D',j} && \text{by definition of } i_{D',j} \\
&\geq a \log i_{D',j}
\end{aligned}$$

for almost every $j$ i.e. $R_{\mathrm{Kt}}$ is monotone-Lin-deep.

## 6 The set of Kolmogorov-random strings is deep

The next result shows that the set of Kolmogorov random strings is bm-Poly-deep.

**Definition 6.1** *Fix a universal Turing machine $U$. The Kolmogorov complexity of $x$ is the length of the shortest program that outputs $x$.*

$$K(x) = \min\{|p|: \ U(p) = x\}.$$

The definition of $K$ does not depend on the choice of the universal TM $U$, up to an additive constant (see [13]).

For a time bound $t$, the $t$-bounded Kolmogorov complexity of $x$ is

$$K^t(x) = \min\{|p|: \ U(p) = x, \text{ and } U \text{ halts in at most } t(|x|) \text{ steps}\}.$$

Let $0 < \epsilon < 1$. The set of Kolmogorov random string is

$$R_{K,\epsilon} = \{x \in \{0,1\}^*: \ K(x) \geq \epsilon|x|\}. \tag{3}$$

**Theorem 6.1** *Let $0 < \epsilon < 1$. $R_{K,\epsilon}$ is a.e. bm-Poly-deep.*

*Proof.* Let $0 < \epsilon < 1$ We need the following lemma.

**Lemma 6.1** *For every Poly-compression $(C, D, p)$ of $R_{K,\epsilon}$ and for almost every $j \in \mathbb{N}$*

$$i_{D,j} < 2^{j+1}.$$

Let us prove the lemma by contradiction. Suppose there is a Poly-compression $(C, D, p)$ of $R_{K,\epsilon}$ and an infinite set $N$ of integers $j$ such that $i_{D,j} \geq 2^{j+1}$. Let $c = 4/(1-\epsilon)$ and $j \in N$.

Let $y_1, \ldots, y_c \in \{0,1\}^j$ such that $K^{2^{n^2}}(\langle y_1, \ldots, y_c \rangle) \geq cj - O(\log j)$ but $K(\langle y_1, \ldots, y_c \rangle) \leq O(\log j)$. Such a $c$-tuple can be found by simulating $U$ on all programs of appropriate size running in at most $2^{n^2}$ steps. We have $R_{K,\epsilon}(y_t) = 0$ for every $t = 1, \ldots, c$.

Consider $L = \{(l_1, \cdots, l_c) \mid 1 \leq l_t \leq 2^{\epsilon(j+1)},\ t = 1, \ldots, c\}$. Let $Q = \{q_l \mid l \in L\}$ with $q_l = \langle \text{code}, p[1..j], l \rangle$ be the set of programs such that $U$ on input $q_l$ simulates $D(p[1..j])$ to reconstruct $R_{K,\epsilon}[1..i_{D,j}] \sqsupseteq R_{K,\epsilon}[1..2^{j+1}]$, which takes time less than $O(2^j)$ ($U$ stops once $D$ already output the $2^{j+1}$ first bits of $R_{K,\epsilon}$). $U$ constructs

$$R_0 = \{r_1 < r_2 < \ldots \mid r_t \in \{0,1\}^{\leq j}, R_{K,\epsilon}(r_t) = 0\}$$

the lexicographical ordered set of all strings of length at most $j$ whose characteristic bit in $R_{K,\epsilon}$ is 0, which takes time $O(2^j)$. If $r_{l_1}, \cdots, r_{l_c} \in R_0$ then output $\langle r_{l_1}, \cdots, r_{l_c} \rangle$ else halt, which takes time $O(2^j)$.

On any program $q_l \in Q$, $U$ runs in less than $2^{O(j)}$ steps. Moreover all $l_t$ ($t = 1, \ldots, c$) can be encoded in at most $\epsilon(j+1)$ bits i.e., all programs $q_l \in Q$ have size bounded by

$$|q_l| \leq c\epsilon(j+1) + j + O(\log j) \leq (c\epsilon + 1)j + O(\log j).$$

Because $R_{K,\epsilon}(y_t) = 0$ for every $t = 1, \ldots, c$, let $v = (v_1, \ldots, v_c) \in L$ be the vector of the positions of $y_1, \ldots, y_c$ in $R_{K,\epsilon}$ i.e., $r_{v_t} = y_t$ for every $t = 1, \ldots, c$. Thus $U$ on input $q_v$ outputs $\langle y_1, \ldots, y_c \rangle$ i.e., $q_v$ is a program for $\langle y_1, \ldots, y_c \rangle$ that runs in less than $2^{O(j)}$ steps. Thus we have

$$K^{2^{n^2}}(\langle y_1, \ldots, y_c \rangle) \leq (c\epsilon + 1)j + O(\log j) \quad \text{which implies}$$
$$cj - O(\log j) \leq (c\epsilon + 1)j + O(\log j) \quad \text{i.e.,}$$
$$cj \leq (c\epsilon + 1)j + O(\log j) \leq (c\epsilon + 2)j$$

thus $c(1-\epsilon) \leq 2$ which is a contradiction. $\square$

**Lemma 6.2** *There exists a Rec-compression $(D, p)$ of $R_{K,\epsilon}$ such that for almost every $j \in \mathbb{N}$*

$$2^{j+1} \geq i_{D,j} \geq 2^{j/\epsilon}.$$

Let $p = \Omega[1..n]$ be the halting probability $\Omega = \sum_{p:U(p)\downarrow} 2^{-|p|}$. $D$ on input $p[1..\epsilon j]$ can compute using standard Dove-tailing (see [13]) whether $U(p) \downarrow$ for all programs $p$ with $|p| \leq \epsilon j$ i.e., it can reconstruct $R_{K,\epsilon}[1..2^{j+1} - 1]$. We have $i_{D,\epsilon j} \geq 2^j$ i.e., $i_{D,j} \geq 2^{j/\epsilon}$. By construction $2^{j+1} \geq i_{D,j}$ $\square$

Let us show that $R_{K,\epsilon}$ is bm-Poly-deep. Let $a > 0$ and $(C, D, p)$ be a Poly-compression of $R_{K,\epsilon}$, and let $(D', p')$ be the Rec-compression from Lemma 6.2. We have

$$\begin{aligned}
i_{D',j} - i_{D,j} &\geq 2^{j/\epsilon} - i_{D,j} && \text{by Lemma 6.2}\\
&\geq 2^{j/\epsilon} - 2^{j+1} && \text{by Lemma 6.1}\\
&= 2^j(2^{(1/\epsilon - 1)j} - 2)\\
&> 2^{j+1} && \text{for } j \text{ large enough}\\
&\geq i_{D',j} && \text{by Lemma 6.2}\\
&\geq a \log i_{D',j}
\end{aligned}$$

for almost every $j$ i.e. $R_{K,\epsilon}$ is a.e. bm-Poly-deep. $\square$

# References

[1] E. Allender, H. Buhrman, and M. Koucký. What can be efficiently reduced to the K-random strings? In V. Diekert and M. Habib, editors, *STACS*, volume 2996 of *Lecture Notes in Computer Science*, pages 584–595. Springer, 2004.

[2] E. Allender, H. Buhrman, and M. Koucký. What can be efficiently reduced to the Kolmogorov-random strings? *Ann. Pure Appl. Logic*, 138(1-3):2–19, 2006.

[3] E. Allender, H. Buhrman, M. Koucký, D. van Melkebeek, and D. Ronneburger. Power from random strings. In *FOCS*, pages 669–678. IEEE Computer Society, 2002.

[4] E. Allender, H. Buhrman, M. Koucký, D. van Melkebeek, and D. Ronneburger. Power from random strings. *SIAM J. Comput.*, 35(6):1467–1493, 2006.

[5] L. Antunes, L. Fortnow, D. van Melkebeek, and N. Vinodchandran. Computational depth: Concept and applications. *Theoretical Computer Science*, 354:391–404, 2006.

[6] K. B. Athreya, J. M. Hitchcock, J. H. Lutz, and E. Mayordomo. Effective strong dimension in algorithmic information and computational complexity. *SIAM J. Comput.*, 37(3):671–705, 2007.

[7] C. H. Bennett. Logical depth and physical complexity. *The Universal Turing Machine, A Half-Century Survey*, pages 227–257, 1988.

[8] H. Buhrman and L. Longpré. Compressibility and resource bounded measure. *SIAM J. Comput.*, 31(3):876–886, 2001.

[9] D. Doty and P. Moser. Feasible depth. In S. B. Cooper, B. Löwe, and A. Sorbi, editors, *CiE*, volume 4497 of *Lecture Notes in Computer Science*, pages 228–237. Springer, 2007.

[10] D. W. Juedes, J. I. Lathrop, and J. H. Lutz. Computational depth and reducibility. *Theor. Comput. Sci.*, 132(2):37–70, 1994.

[11] J. I. Lathrop and J. H. Lutz. Recursive computational depth. *Inf. Comput.*, 153(1):139–172, 1999.

[12] L. A. Levin. Randomness conservation inequalities; information and independence in mathematical theories. *Information and Control*, 61(1):15–37, 1984.

[13] M. Li and P. Vitanyi. *Introduction to Kolmogorov complexity and its applications*. Springer, 1993.

[14] J. Lutz. Almost everywhere high nonuniform complexity. *Journal of Computer and System Science*, 44:220–258, 1992.

[15] P. Moser. A general notion of useful information. In T. Neary, D. Woods, A. K. Seda, and N. Murphy, editors, *CSP*, volume 1 of *EPTCS*, pages 164–171, 2008.

[16] P. Moser. Martingale families and dimension in P. *Theor. Comput. Sci.*, 400(1-3):46–61, 2008.

# A  Appendix

*Proof.* (of Thm 4.1) Let $S$ be a Poly-optimally-compressible sequence, and let $(C_0, D_0, p_0)$ be a Poly-compressor witnessing this fact. By definition, $S$ is Poly-shallow if there exists a Poly-compression $(C, D, p)$ of $S$ and $a > 0$, such that for any Poly-compression $(C', D', p')$ of $S$, and for infinitely many $j \in \mathbb{N}$

$$i_{D',j} - i_{D,j} < a \log i_{D',j}.$$

Letting $a = 1$ and $(C, D, p) = (C_0, D_0, p_0)$ implies that for almost every $j \in \mathbb{N}$, $i_{D,j} = \mathrm{MD}(j)$. Let $(C', D', p')$ be any Poly-compression of $S$, then for almost every $j \in \mathbb{N}$

$$i_{D',j} - i_{D,j} \leq \mathrm{MD}(j) - i_{D,j} = \mathrm{MD}(j) - \mathrm{MD}(j) = 0 < \log i_{D',j}.$$

□

*Proof.* (of Theorem 4.2) Let $S$ be a Poly-random sequence, and let $(C, D, p)$ be the identity compressor i.e. $p = S$ and $C(x) = D(x) = x$ for any string $x$ (in particular $i_{D,j} = j$). Let $a = 1$. Let us show that for any Poly-compression $(C', D', p')$ of $S$, and for infinitely many $j \in \mathbb{N}$

$$i_{D',j} - i_{D,j} < a \log i_{D',j}.$$

Let $(C', D', p')$ be any Poly-compression of $S$; then there exists $c \in \mathbb{N}$ such that for almost every $j \in \mathbb{N}$, $i_{D',j} - j \leq c$. Thus

$$i_{D',j} - i_{D,j} = i_{D',j} - j \leq c < \log i_{D',j}$$

for almost every $j \in \mathbb{N}$. □

*Proof.* (of Theorem 4.3) Let $S, T$ be as above, and let $M$ denote the Poly-monotone reduction. We need the following result that shows that $M$ can be inverted in polynomial time.

**Lemma A.1** *Let $M$ be as above. Then there is a polynomial TM $N$ such that for every string $x$, $N(M(x)) = x'$, with $x' \sqsupseteq x$, and $M(x') = M(x)$ (and $x'$ is maximal i.e., $\forall y \in \{0,1\}^*$, $M(x'y) \neq M(x')$).*

Let us prove the lemma. Let $M$ be as above, let $x, y \in \{0,1\}^*$ such that $M(x) = y$, $|x| = n$, $|y| = m$. Let $c$ be given by honesty of $M$, thus

$$n - c \log n \leq |y| \leq n + c \log n \quad \text{which implies} \tag{4}$$

$$|y| - c \log n \leq n \leq |y| + c \log n \quad \text{and}$$

$$\log \frac{n}{2} \leq \log(n - c \log n) \leq \log |y| \quad \text{i.e.}$$

$$\log n \leq 2 \log |y|.$$

Therefore

$$|y| - 2c \log |y| \leq n \leq |y| + 2c \log |y|. \tag{5}$$

String $x$ is constructed recursively, by groups of $\log m$ bits at a time, i.e we construct $x_1 \sqsubset x_2 \sqsubset \ldots$ such that the final $x_f$ will be the promised $x'$. The construction is a follows, suppose $x_1 \sqsubset x_2 \sqsubset \ldots \sqsubset x_i$ have already been constructed. We construct $x_{i+1}$ as follows; try all $m$ strings $z \in \{0,1\}^{\log m}$ to find the unique $z$ such that $M(x_i z) \sqsubset y$, and let $x_{i+1} = x_i z$. $z$ is unique; suppose by contradiction that there are $z, z'$ such that $M(x_i z) \sqsubset y$ and $M(x_i z') \sqsubset y$ i.e. $M(x_i z) \sim M(x_i z')$, thus by monotone injectivity, $x_i z \sim x_i z'$, implying $z = z'$ (since $|z| = |z'|$). The procedure is carried on until $x_{f-1}$ with $|x_{f-1}| = |y| - 2c \log |y|$. Finally all extensions $z \in \{0,1\}^{\leq 4c \log |y|}$ are tested to find the longest such $z$ such that $M(x_{f-1} z) = y$. By Equation 5 such a $z$ exists, and letting $x' = x_{f-1} z$, guarantees that $x'$ is maximal. As both the number of extensions to test during the whole constructions of $x'$ and the number of steps $1, \ldots, f$ is polynomial in $|y|$, $N$ runs in polynomial time; which ends the proof of the lemma. □

Let us show that $T$ is Poly-deep; let $a > 0$ and $(C, D, p)$ be a Poly-compression of $T$. This induces a Poly-compression $(C_1, D_1, p)$ for $S$, with $D_1 = M \circ D$ and $C_1 = C \circ N$. $C_1, D_1$ run in Poly-time. For any $j, n \in \mathbb{N}$ we have

$$D_1(p[1..j]) = M(T[1..i_{D,j}]) \sqsupseteq S[1..i_{D,j} - c \log i_{D,j}]$$

by Equation 4, which implies

$$i_{D_1, j} \geq i_{D,j} - c \log i_{D,j}. \tag{6}$$

Also

$$C_1(S[1..n]) = C \circ N(S[1..n]) = C(T[1..m])$$

with $M(T[1..m]) = S[1..n]$, and $C(T[1..m])$ outputs several strings one of which is a prefix $x$ of $p$, such that

$$T \sqsupseteq D(x) \sqsupseteq T[1..m].$$

Thus

$$D_1(x) = M(D(x)) \sqsupseteq M(T[1..m]) = S[1..n].$$

Let $b = 2a + 6c > 0$. Since $S$ is Poly-deep, there exists a Poly-compression $(C_2, D_2, p_2)$ for $S$ such that for every $j \in \mathbb{N}$

$$i_{D_2, j} - i_{D_1, j} > b \log i_{D_2, j}. \tag{7}$$

We construct a Poly-compression $(C', D', p_2)$ of $T$ with $D' = N \circ D_2$ and $C' = C_2 \circ M$. $C', D'$ run in Poly-time. For any $j, n \in \mathbb{N}$ we have

$$D'(p_2[1..j]) = N(S[1..i_{D_2, j}]) \sqsupseteq T[1..i_{D_2, j} - 2c \log i_{D_2, j}]$$

implying

$$i_{D', j} \geq i_{D_2, j} - 2c \log i_{D_2, j} \tag{8}$$

and

$$C'(T[1..n]) = C_2(M(T[1..n])) = C_2(S[1..n'])$$

where $n'$ is an integer such that $n' \geq n - c \log n$, $N(S[1..n']) \sqsupseteq T[1..n]$ and $C_2(S[1..n'])$ outputs several strings one of which is a prefix $x'$ of $p_2$, such that

$$S \sqsupseteq D_2(x') \sqsupseteq S[1..n'].$$

Thus

$$D'(x') \sqsupseteq N(S[1..n']) \sqsupseteq T[1..n].$$

By Equation 6 we have

$$\log i_{D_1,j} \geq \log(i_{D,j} - c\log i_{D,j}) \geq \frac{1}{2}\log i_{D,j}. \qquad (9)$$

Similarly Equation 7 yield

$$\log i_{D_1,j} \leq 2\log i_{D_2,j} \qquad (10)$$

and since $D' = N \circ D$

$$\log i_{D',j} \leq 2\log i_{D_2,j}. \qquad (11)$$

Thus for almost every $j$

$$\begin{aligned}
i_{D',j} - i_{D,j} &\geq i_{D_2,j} - i_{D,j} - 2c\log i_{D_2,j} & \text{by Equation 8} \\
&\geq i_{D_2,j} - i_{D_1,j} - 2c\log i_{D_2,j} - c\log i_{D,j} & \text{by Equation 6} \\
&\geq i_{D_2,j} - i_{D_1,j} - 2c\log i_{D_2,j} - 2c\log i_{D_1,j} & \text{by Equation 9} \\
&\geq i_{D_2,j} - i_{D_1,j} - 2c\log i_{D_2,j} - 4c\log i_{D_2,j} & \text{by Equation 10} \\
&\geq (b - 6c)\log i_{D_2,j} & \text{by depth of } S \\
&\geq (b - 6c)/2 \log i_{D',j} & \text{by Equation 11} \\
&\geq a\log i_{D',j} & \text{by definition of } b
\end{aligned}$$

i.e. $T$ is deep. □

*Proof.* (of Theorem 4.4) As shown in [8] polynomial compression yields an alternative characterization of resource-bounded measure zero sets. Resource-bounded measure is a measure theory within the complexity class E developed by Lutz [14], which is obtained by imposing polynomial resource-bounds on a game theoretical characterization of classical Lebesgue measure, via martingales. A martingale is a function $d : \{0,1\}^* \to \mathbb{R}_+$ such that, for every $w \in \{0,1\}^*$, $2d(w) = d(w0) + d(w1)$. This definition can be motivated by the following betting game in which a gambler puts bets on the successive membership bits of a hidden language $A$. The game proceeds in infinitely many rounds where at the end of round $n$, it is revealed to the gambler whether $s_n \in A$ or not. A polynomial (computable) martingale is a martingale computable in time polynomial in the input size.

In [8] the following equivalence between polynomial martingales and monotone compression was shown.

**Lemma A.2** *Given a polynomial computable martingale $d$, and a sequence $w$, there exists a Poly-compression $(C, D)$ for $w$ such that for any $j \in \mathbb{N}$*

$$i_{D,j} - j \geq \log d(w[1..i_{D,j}]) - 4.$$

*Alternatively given a Poly-compression $(C, D)$ for $w$, there exists a polynomial martingale $d$ such that for any $j \in \mathbb{N}$*

$$\log d(w[1..i_{D,j}]) \geq i_{D,j} - j - 2.$$

It was shown in [14] that for every $k \in \mathbb{N}$ there exists a $n^k$-universal martingale $d_k$ computable in polynomial time, such that for any martingale $d$ computable in time $n^k$, there exists $c > 0$, such that for any $w \in \{0,1\}$

$$d_k(w) \geq c \cdot d(w).$$

The sequence $S = S_1^1 S_1^2 S_2^2 \ldots S_1^l \ldots S_l^l$ is constructed by induction, where $|S_1^1| = 1$ and $|S_k^l| = \text{MD}(\text{MD}(|S \upharpoonright S_k^l|))$, and $S_k^l$ diagonalizes against $d_k$ (i.e. $d_k$ does not increase on $S_k^l$); more precisely, define $S_k^l$ by induction, where for every $t \in \{0, \ldots, |S_k^l| - 1\}$, the next bit $b \in \{0, 1\}$ of $S_k^l$ is chosen such that

$$d_k((S \upharpoonright S_k^l)S_k^l[0..t]b) \leq d_k((S \upharpoonright S_k^l)S_k^l[0..t]).$$

We need the following lemma.

**Lemma A.3** *For any $k, l \in \mathbb{N}$ ($k \leq l$), there exists $j = j(k, l) \in \mathbb{N}$ such that for any Poly-compression $(C, D)$ for $S$*

$$3|S \upharpoonright S_k^l| \leq i_{D,j} \leq |S \upharpoonright S_k^l|.$$

Let us prove the lemma, wlog (by ignoring the decompressors that do not output at least $n$ bits from a program of size $2n$) for any Poly-compression $(C, D)$ and for almost every $j \in \mathbb{N}$,

$$j/2 \leq i_{D,j} \leq \text{MD}(j).$$

Letting $b = |S \upharpoonright S_k^l|$ and $j = \text{MD}(b)$ yields

$$i_{D,j} \geq j/2 = \text{MD}(b)/2 \geq 3b$$

and

$$i_{D,j} \leq \text{MD}(j) = \text{MD}(\text{MD}(b)) = |S_k^l| \leq |S \upharpoonright S_k^l|$$

which ends the proof of the lemma.

Let us show that $S$ is monotone-Poly-deep. Let $(C, D)$ be a Poly-compression of $S$, and let $a > 0$. By Lemma A.2, let $d$ be a Poly martingale s.t. for any $j \in \mathbb{N}$

$$\log d(w[1..i_{D,j}]) \geq i_{D,j} - j - 2$$

and suppose $d$ runs in time $n^k$. Thus there is a $c > 0$ such that $cd_k(w) \geq d(w)$ for any string $w$. Let $j$ be given by Lemma A.3. By definition of $S_k^l$, $d_k$ does not increase on it; as $d_k$ can at most double its capital on every bit of $S \upharpoonright S_k^l$, and by Lemma A.3 we have

$$i_{D,j} - j - O(1) \leq \log d_k(S[1..i_{D,j}]) \leq |S \upharpoonright S_k^l| \leq i_{D,j}/3$$

i.e.

$$i_{D,j} \leq 3j/2 + O(1)$$

where the constant $O(1)$ does not depend on $j$.

Consider the following Poly-compression $(C', D', p)$ for $S$. Informally $D'$ reconstructs $S_k^l$ using $d_k$. Program $p$ is equal to $S$, except on blocks $S_k^l$ (for every $l \in \mathbb{N}$), where $p$ is $j' = j(k, l) - |p \upharpoonright S_k^l|$ zeroes, i.e more formally

$$p = S_1^1 S_1^2 S_2^2 \ldots S_1^l \ldots S_{k-1}^l 0^{j'(k,l)} S_{k+1}^l \ldots S_l^l.$$

Since it is easy to compute the sizes of the blocks $S_k^l$, it is easy to determine where each block starts and stops in $p$. $D'$ on input a prefix of $p'$ of $p$ with $|p'| = j(k, l)$, can reconstruct the parts in $S$ not in an $S_k^l$ ($l \in \mathbb{N}$) block, by just reading $p'$. The ending of $p'$ (which corresponds to an $S_k^l$ block), consists of $j'(k, l)$ zeroes. On every such zeroes except the last one, $D'$ outputs one bit of $S_k^j$ (reconstructed using $d_k$). On the last zero, $D'$ outputs $\mathrm{MD}(j) - j' + 1$ bits of $S_k^l$. After the last zero, $D'$ will have output $\mathrm{MD}(j)$ bits of $S_k^l$, i.e. will have output exactly all bits of $S_k^l$, and we have

$$i_{D',j} = \mathrm{MD}(j).$$

The definition of $C'$ is trivial and left to the reader. We have

$$\begin{aligned} i_{D',j} - i_{D,j} &= \mathrm{MD}(j) - i_{D,j} \\ &\geq \mathrm{MD}(j) - 3j/2 - O(1) \\ &\geq a \log \mathrm{MD}(j) \\ &\geq a \log i_{D',j}. \end{aligned}$$

Since there are infinitely many $j = j(l, k)$ (one for every $l \geq k$), $S$ is deep. □